\documentclass[runningheads]{llncs}
\usepackage[T1]{fontenc}
\usepackage{graphicx}
\usepackage{enumitem} 
\usepackage[shortcuts,acronym]{glossaries}
\usepackage{xcolor}
\usepackage{algorithm}
\usepackage{algpseudocode}
\usepackage{xspace}
\usepackage[disable]{todonotes}
\usepackage{amsfonts}
\usepackage{tikz}
\usepackage{booktabs}
\usepackage{colortbl}
\usepackage{hyperref}
\usepackage{tcolorbox}
\usepackage{url}
\usepackage{subcaption}

\graphicspath{{images}{.pdf}}
\newcommand*\circled[1]{%
  \tikz[baseline=(char.base)]{%
    \node[shape=circle, draw, inner sep=0.1pt, 
          fill=black, text=white, font=\bfseries] (char) {#1};}%
}

\usepackage{enumitem} 
\newcommand{\packhero}{PackHero\xspace}
\newcommand{\yara}{YARA\xspace}

\newcommand{\mypar}[1]{\smallskip\noindent\textbf{#1.}\xspace}

\begin{document}
\title{\packhero: A Scalable Graph-based Approach for Efficient Packer Identification}
\titlerunning{\packhero}
%
\author{Marco Di Gennaro \and
Mario D'Onghia \and
Mario Polino \and
Stefano Zanero \and
Michele Carminati}

\authorrunning{M. Di Gennaro et al.}

\institute{Dipartimento di Elettronica, Informazione e Bioingegneria,\\ Politecnico di Milano, Milan, Italy \\
\email{\{marco.digennaro,mario.donghia,mario.polino,\\stefano.zanero,michele.carminati\}@polimi.it}}
\newacronym{ml}{ML}{Machine Learning}
\newacronym{ai}{AI}{Artificial Intelligence}
\newacronym{dl}{DL}{Deep Learning}
\newacronym{gml}{GML}{Graph Machine Learning}
\newacronym{av}{AV}{Anti-Virus}
\newacronym{avs}{AVs}{Anti-Viruses}
\newacronym{sota}{SotA}{State-of-the-art}
\newacronym{gnn}{GNN}{Graph Neural Network}
\newacronym{nn}{NN}{Neural Network}
\newacronym{gnns}{GNNs}{Graph Neural Networks}
\newacronym{gmn}{GMN}{Graph Matching Network}
\newacronym{gmns}{GMNs}{Graph Matching Networks}
\newacronym{gcn}{GCN}{Graph Convolutional Network}
\newacronym{gat}{GAT}{Graph Attention Network}
\newacronym{pe}{PE}{Portable Executable}
\newacronym{pes}{PEs}{Portable Executables}
\newacronym{ged}{GED}{Graph Edit Distance}
\newacronym{mcs}{MCS}{Maximum Common Subgraph}
\newacronym{tp}{TP}{True Positive}
\newacronym{fp}{FP}{False Positive}
\newacronym{fps}{FPs}{False Positives}
\newacronym{tn}{TN}{True Negative}
\newacronym{fn}{FN}{False Negative}
\newacronym{fns}{FNs}{False Negatives}
\newacronym{fpr}{FPR}{False Positive Rate}
\newacronym{tpr}{TPR}{True Positive Rate}
\newacronym{mlp}{MLP}{Multilayer Perceptron}
\newacronym{cfg}{CFG}{Control-Flow Graph}
\newacronym{cfgs}{CFGs}{Control-Flow Graphs}
\newacronym{ddg}{DDG}{Data Dependence Graph}
\newacronym{ddgs}{DDGs}{Data Dependence Graphs}
\newacronym{cg}{CG}{Call Graph}
\newacronym{cgs}{CGs}{Call Graphs}
\newacronym{vm}{VM}{Virtual Machine}
\newacronym{db}{DB}{Database}
\newacronym{os}{OS}{Operating System}
\newacronym{auc}{AUC}{Area Under the ROC Curve}
\newacronym{roc}{ROC}{Receiver Operating Characteristic}
\newacronym{cpu}{CPU}{Central Processing Unit}
\newacronym{die}{DIE}{Detect It Easy}
\newacronym{rq}{RQ}{Research Question}
\newacronym{rqs}{RQs}{Research Questions}
\newacronym{tdr}{TDR}{Total Detection Rate}
\newacronym{ceg}{CEG}{Consistently-Executing Graph}
\newacronym{cegs}{CEGs}{Consistently-Executing Graphs}

\maketitle              
\begin{abstract}
    \label{sec:abstract}

Anti-analysis techniques, particularly packing, challenge malware analysts, making packer identification fundamental. Existing packer identifiers have significant limitations: signature-based methods lack flexibility and struggle against dynamic evasion, while \acrlong{ml} approaches require extensive training data, limiting scalability and adaptability. Consequently, achieving accurate and adaptable packer identification remains an open problem. This paper presents \packhero, a scalable and efficient methodology for identifying packers using a novel static approach. \packhero employs a \acrlong{gmn} and clustering to match and group \acrlong{cgs} from programs packed with known packers. 
We evaluate our approach on a public dataset of malware and benign samples packed with various packers, demonstrating its effectiveness and scalability across varying sample sizes. \packhero achieves a macro-average F1-score of 93.7\% with just 10 samples per packer, improving to 98.3\% with 100 samples. Notably, \packhero requires fewer samples to achieve stable performance compared to other \acrlong{ml}-based tools. Overall, \packhero matches the performance of \acrlong{sota} signature-based tools, outperforming them in handling Virtualization-based packers such as Themida/Winlicense, with a recall of 100\%.

\end{abstract}

\section{Introduction}\label{sec:intro}

\textit{Code packing}, a widely used \textit{anti-analysis technique}~\cite{galloro_systematical_2022}, affects the performance of both \acrfull{ml}-based and traditional signature-based malware detection systems~\cite{low_entropy_analysis}. Packers encrypt or compress executable code, rendering static analysis ineffective~\cite{ugarte-pedrero_sok_2015}. At runtime, an \textit{unpacking stub} embedded in the executable restores the original code by decrypting or decompressing it in memory, allowing the program to execute.
This ability to bypass static analysis makes packing particularly appealing to malware authors. The prevalence of \textit{packed} malware can bias ML-based detectors into flagging all packed executables as malicious~\cite{aghakhani_when_2020}. However, this assumption is flawed, as many benign programs (\textit{goodware}) are also packed to protect intellectual property. For instance, Rahbarinia et al.~\cite{rahbarinia_exploring_2017} found that 54\% of benign programs and 58\% of malware samples in their study were processed with known packers, demonstrating a similar distribution.

Recovering packed code might seem feasible through \textit{dynamic analysis}, as executing a packed program can trigger it to unpack itself. However, modern malware increasingly employs dynamic \textit{evasive behaviors} designed to detect analysis environments and prevent the program from exposing its true functionality at runtime~\cite{galloro_systematical_2022}. As a result, packed malware with evasive tactics often resists dynamic-based analysis techniques. Additionally, incorporating dynamic analysis mechanisms into commercial \acrshort{avs} presents challenges, such as requiring kernel-level privileges to execute untrusted code~\cite{survey_anti_analysis,aghakhani_when_2020} and introducing significant computational overhead due to the virtualization infrastructure~\cite{low_entropy_analysis}. Alternatively, \textit{static} identification of the specific packer used in a malware sample could allow \acrshort{avs} to retrieve the original code, if possible, by executing a corresponding unpacker when available. Previous works in this area have applied signature-based methods~\cite{detectiteasy} or \acrshort{ml}-based algorithms using static features~\cite{binary_diffing,liu_2-spiff_2021,sun_pattern_2010}. While effective for known packers, these approaches demand substantial effort to accommodate new packers or variations of existing ones. This challenge is amplified by the frequent emergence of \textit{custom packers} in novel malware~\cite{sun_pattern_2010}, necessitating either extensive manual signature analysis (e.g., with \textit{Detect It Easy}~\cite{detectiteasy}) or complete re-training of \acrshort{ml}-based models.
Recent work proposed \textit{PackGenome}, a tool that automates YARA rule generation from packed samples to detect packed binaries~\cite{li_packgenome_2023}. While effective on large and heterogeneous datasets, it relies on dynamic analysis, requiring packers to generate custom-packed samples, limiting the integration of newly discovered packers.
These limitations motivated our research into new methodologies for code packer identification, focusing on minimizing packer integration effort. The primary challenge lies in achieving a balance between \textit{accuracy}, rapid \textit{adaptability} for integrating newly discovered packers, resilience against dynamic \textit{evasion} techniques, and overall \textit{scalability}.

This paper presents \packhero, a packer identifier that leverages statically extracted \acrlong{cg}s from packed programs. \packhero extracts the \acrshort{cg} of a given binary and identifies the packer by comparing it with previously labeled graphs in a stored collection. The graph representation is inspired by the work of X. Li et al.~\cite{li_consistently-executing_2019}. \acrshort{cgs} enable a high level of abstraction and reveal that portions of these graphs remain identical or similar for binaries packed with the same packer. To leverage this, we introduce a heuristic to isolate the graph segment corresponding to the unpacking stub, identifying unique patterns shared by binaries processed by the same packer. To solve the graph similarity problem, we use a specific \acrfull{gnn}~\cite{hamilton_graph_2020}, known as a \acrfull{gmn}~\cite{li_graph_2019}. Additionally, \packhero incorporates a hierarchical clustering approach to group similar graphs, enhancing identification accuracy while ensuring constant inference time when integrating new packers.

We evaluate \packhero on a publicly available dataset of packed Windows \acrfull{pes}~\cite{aghakhani_when_2020}, containing both malware and benign samples, repacked with various commercial and free packers, categorized by complexity according to existing taxonomies~\cite{ugarte-pedrero_sok_2015}. 
\packhero achieves a macro-average F1-score of 93.7\% and an accuracy of 98.7\% using only 10 programs per packer during configuration. In its best configuration, utilizing 100 samples per packer, it reaches a macro-average F1-score of 98.3\% and an accuracy of 99.8\%. The scalability of \packhero, supported by its clustering approach, is validated through comparisons with a non-clustering version in terms of both performance and inference calls to the \acrshort{gmn}. Once configured, \packhero requires significantly fewer samples to converge and stabilize than existing \acrshort{ml}-based tools, needing just 10 samples versus 40 for the best-performing alternative. Moreover, \packhero features a constant integration cost, whereas the integration cost of other \acrshort{ml}-based tools increases linearly with the number of packers recognized. \packhero is a robust alternative to signature-based detection tools, achieving performance aligned with \acrshort{sota} tools. Notably, it performs significantly better against virtualization-based~\cite{virtbasedpackers} packers like Themida/Winlicense, which employ advanced dynamic evasive behaviors that hinder signature extraction in dynamic analysis-based tools. Specifically, \packhero achieves a perfect recall of 100\% on this packer, compared to 92\% for \acrshort{die} and 31\% for PackGenome.

In summary, the contributions are the following :
\begin{itemize}
    \item A hybrid \acrshort{ml} and graph signature-based approach for packer identification, enabling automatic and scalable integration of both accessible and non-accessible packers (e.g., custom or closed-source) directly from packed programs throughout the tool lifecycle.
    \item A heuristic to statically extract a \acrlong{cg} of an unpacking routine or a part of it from a packed binary.
    \item A combination of \acrfull{gmn} with hierarchical clustering to enhance the accuracy and reduce the search space of similar graphs.
    \item We release \packhero's source code\footnote{\url{https://github.com/necst/packhero}} for reproducibility.
\end{itemize}
\section{Background and Motivation}
\label{sec:rel_work}

\textit{Code packing} is a widely used anti-analysis technique~\cite{file_packing_survey}, where packers, acting as third-party software, transform a program's structure and content, recovering the original software at runtime via a \textit{tail jump} to the original entry point. Initially intended for file compression, most packers now aim to obfuscate and hinder program analysis in legitimate and malicious software. Packers are classified by runtime complexity~\cite{ugarte-pedrero_sok_2015} into six types (I–VI), with most common packers falling within types I–III. Another taxonomy focuses on obfuscation methods, distinguishing \textit{compressors}, \textit{crypters}, and \textit{virtualization-based} packers, such as Themida~\cite{virtbasedpackers}, which translate code into virtual instructions and implement advanced anti-dynamic analysis techniques. In our experiments, we consider packers from types I–III, with Themida representing the \acrshort{vm}-based category.

\subsection{Packer Identification}

\textit{Packer identification} is a multi-label classification task aimed at determining the specific packer used to compress or obfuscate a program. This capability allows \acrshort{av} tools to statically unpack programs, thereby enhancing malware detection~\cite{low_entropy_analysis}. In contrast, \textit{packer detection} identifies whether a program is packed, often employing static methods such as similarity comparisons~\cite{packed_similarity,detection_packed_exes} or entropy analysis~\cite{entropy_1,entropy_2}. However, these methods are less effective against \textit{low-entropy} packers~\cite{low_entropy_analysis}. This paper focuses on packer identification and categorizes existing approaches into two main families: signature-based methods, which rely on manually or automatically generated signatures, and pattern recognition techniques, predominantly driven by \acrshort{ml}-based algorithms.

\mypar{Signature-based Methods}
\label{subsec:signature_background}
Packers often leave specific artifacts that can be used to create signature databases. \textit{\acrfull{die}} is a well-known tool for packer identification via signature matching~\cite{detectiteasy}, outperforming tools like PEiD~\cite{peid} with its open architecture that allows users to add JavaScript-like scripts for packer detection. However, it requires the manual creation of signature scripts for new packers and their variants, making it challenging to integrate new packers, especially with limited analyzable packed samples. A key limitation of signature-based detection is the need to analyze many samples to identify invariant byte sequences that can be used as signatures.
To address this, researchers have explored automating the signature extraction process. Raff et al. propose a method to automatically generate \yara rules~\cite{autoyara}, a format for defining malware characteristics~\cite{yara}. Nevertheless, code packing can still easily defeat these rules, similar to other signature schemes.
To the best of our knowledge, the \acrfull{sota} tool for signature generation in packed programs is PackGenome~\cite{li_packgenome_2023}. Inspired by biological processes, PackGenome identifies significant instructions in the first unpacking layer (the only statically visible one). It uses Intel \textit{Pintool}~\cite{intelpin} to monitor packed programs in a controlled environment, recording and labeling instructions that write ``unpacked'' instructions. By analyzing multiple executions of programs packed with the same packer and applying similarity metrics, PackGenome extracts \textit{packer-specific} ``genes'' to generate \yara rules. However, this approach relies on dynamic analysis, making it vulnerable to evasive techniques that hinder the extraction of relevant genes, as empirically confirmed in our experimental evaluation (Subsection~\ref{subsec:signaturebased_comparison}). Accurate packer identification often requires generating a large number of signatures. For instance, PackGenome recommends using the actual packer to create extensive variations of the unpacking stub. However, this approach is impractical in real-world scenarios where malicious software frequently employs custom packers that are inaccessible to analysts. Additionally, the limited availability of samples for such packers makes it infeasible to build a comprehensive signature database.

\mypar{ML-based Packer Identification}
The second family of identification methods relies on pattern recognition algorithms, particularly \acrlong{ml} techniques. Proposed approaches include constructing randomness profiles for packed samples~\cite{sun_pattern_2010}, applying binary diffing~\cite{binary_diffing}, extracting features from the topology of \acrshort{cgs}~\cite{liu_2-spiff_2021}, and evaluating the similarity of \acrfull{cegs}~\cite{li_consistently-executing_2019}.
S. Li et al. observed that while most packers significantly affect binary entropy, individual packers exhibit distinctive randomness patterns~\cite{sun_pattern_2010}. They used sliding windows~\cite{sliding_windows} to build randomness profiles, training a k-nearest neighbor classifier.
Kim et al. employed an SVM classifier with binary diffing measures as kernels, achieving the best performance using the \textit{longest common substring} computed from the first 15 bytes at each program's entry point~\cite{binary_diffing}. This method leverages the similarity of initial instructions in unpacking stubs from the same packer, but its effectiveness is lowered by code obfuscation~\cite{li_packgenome_2023,symbolic_execution}.
Hao et al. represented packed programs as \acrshort{cgs} and trained an SVM classifier using topological features (e.g., entry point indegree) and general file information like size or section count~\cite{liu_2-spiff_2021}. While we draw on this idea to represent packed programs through \acrshort{cgs}, our methodology directly uses graphs, offering better generalization and results. 
X. Li et al.~\cite{li_consistently-executing_2019} proposed a similar approach by comparing graphs using a Weisfeler-Lehman shortest path kernel. Instead of employing \acrshort{cgs}, they introduce \acrshort{cegs} to simulate static execution points by traversing procedures, locating branch instructions, and forming flow paths. However, they do not address the challenge posed by the increasing number of graphs that must be compared during identification due to the introduction of new packers. To the best of our knowledge, none of these works have released their code. However, three of the four approaches were straightforward to implement, enabling their comparison with our method (results in Subsection~\ref{subsec:integration_new_packers}). The fourth, \acrshort{ceg}, relies on a heuristic for graph extraction, making reimplementation challenging without significant assumptions. Therefore, it was excluded from our study.

\subsection{Motivation}
\label{sec:requirements}

\begin{table}[bt]
\caption{Packer identifiers and compliance with requirements R1–R4.}
\label{tab:sota}
\centering
\resizebox{0.75\textwidth}{!}{
\begin{tabular}{l|cccc|cccccc}
\toprule
\textbf{Work} & \textbf{Analysis Type} & \textbf{Approach} & \textbf{Code} & \textbf{Replicable} & \textbf{R1} & \textbf{R2} & \textbf{R3} & \textbf{R4} \\
\midrule
PackGenome~\cite{li_packgenome_2023} & Static + Dynamic & Signature & \checkmark & \checkmark & \checkmark & \checkmark & & \checkmark \\
Detect It Easy (DIE)~\cite{detectiteasy} & Static & Signature & \checkmark & \checkmark & \checkmark & & \checkmark & \checkmark \\
Randomness Profiles~\cite{sun_pattern_2010} & Static & ML & & \checkmark & \checkmark & \checkmark & \checkmark & \\
Binary Diffing~\cite{binary_diffing} & Static & ML & & \checkmark & \checkmark & \checkmark & \checkmark & \\
2SPIFF~\cite{liu_2-spiff_2021} & Static & ML & & \checkmark & \checkmark & \checkmark & \checkmark & \\
\acrshort{ceg}~\cite{li_consistently-executing_2019} & Static & Signature & & & \checkmark & \checkmark & \checkmark & \\
\midrule
\textbf{\packhero}~(our approach) & Static & ML + Signature & \checkmark & \checkmark & \checkmark & \checkmark & \checkmark & \checkmark\\
\bottomrule
\end{tabular}}
\vspace{-0.5cm}
\end{table}

Given the limitations of current works, we define key requirements for a novel packer identifier. Table~\ref{tab:sota} shows that existing \acrshort{sota} approaches only partially meet these requirements, highlighting the need for our solution.

\mypar{Requirement 1 - High Identification Accuracy} It must achieve high accuracy and low false positives across diverse packer types.

\mypar{Requirement 2 - Efficient Packer Integration} It must efficiently integrate packers using a limited number of real-world samples, addressing challenges such as inaccessible packers and variations within a single packer family. It should rapidly update its identification capabilities without requiring extensive or frequent retraining. To achieve this, the system should leverage robust algorithms that generalize from existing data, enabling the detection of new variations by integrating them using limited samples with minimal manual intervention.

\mypar{Requirement 3 - Dynamic Evasive Behavior Management} It must effectively handle evasive behaviors encountered during the collection of wild samples for integration. Wild samples may employ dynamic evasion techniques or be so damaged or corrupted that execution is impossible~\cite{dailymalware}, thereby complicating dynamic signature extraction performed by state-of-the-art tools~\cite{li_packgenome_2023}. To address this challenge, a static analysis-based approach can mitigate these issues by extracting valuable information and artifacts independently of execution.

\mypar{Requirement 4 - Scalability} It must handle a growing number of packers without performance degradation, integrating new ones seamlessly without major architectural changes or resource demands. In other words, it must handle many/several packer families in parallel.

\section{\packhero}\label{sec:packhero}

\packhero is a \textit{packer identifier} that determines the specific packer used for a given \textit{packed} program. Its approach mirrors the workflow of signature-based detection mechanisms but uses graph ``signature'' to represent packed programs, with matches determined by similarity rather than exact matching.
\packhero leverages a specialized \acrfull{gnn} called \acrfull{gmn}~\cite{li_graph_2019}. It operates on \textit{\acrfull{cgs}} extracted using heuristics. \acrshort{cgs}, which represent the invocation relationships between functions in an executable program~\cite{call_graphs}, are chosen for their compact structure and high level of abstraction. Compared to other binary graph representations (e.g., \acrfull{cfg} and \acrfull{ddg}), \acrshort{cgs} enable an efficient resolution of the similarity problem with \acrshort{gmns}. We divide our approach
into two main phases: \textit{configuration} and \textit{inference}.

\begin{figure}[tb]
  \centering
  \includegraphics[width=0.9\linewidth]{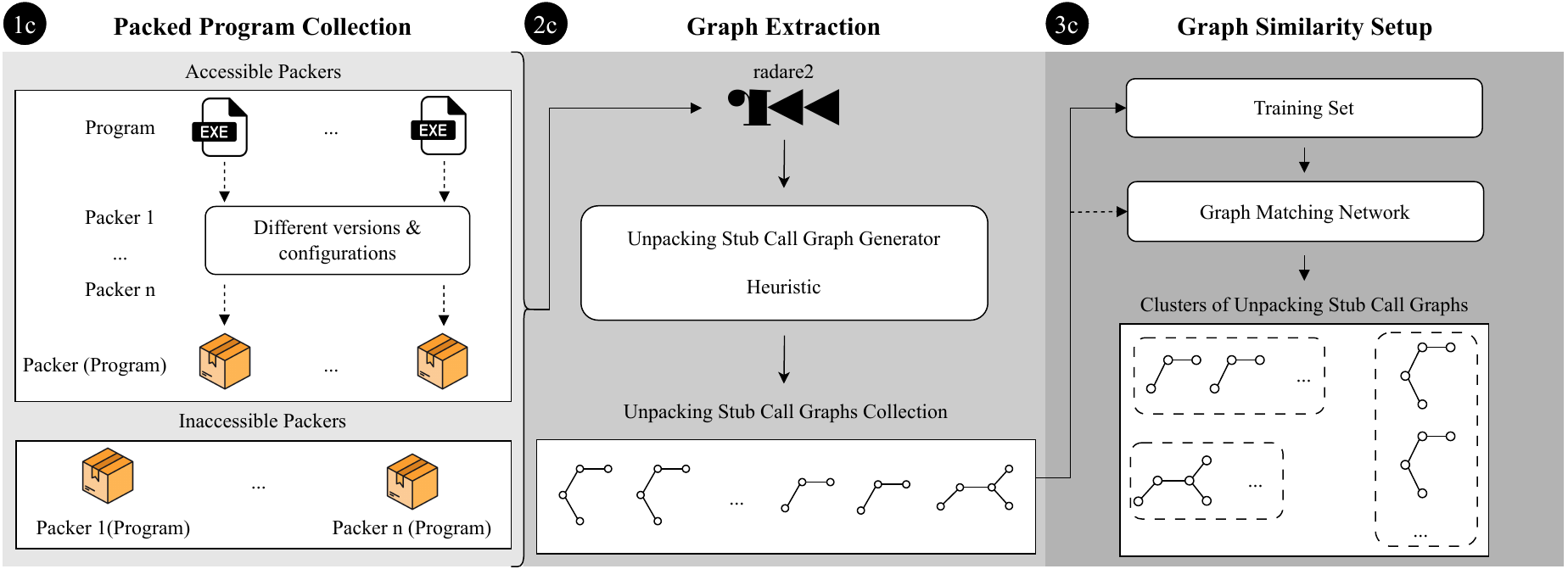}
  \caption{\packhero~Configuration Workflow.}
    \label{fig:config_workflow}
    \vspace{-0.5cm}
\end{figure}

\subsection{Configuration}\label{subsec:configuration}

 As depicted in Fig.~\ref{fig:config_workflow}, this phase involves three main step.

\noindent \circled{1c} \textbf{Collecting Packed Programs.} The first step consists of collecting programs for the packers we want to integrate into the tool. It is important to distinguish between an \textit{accessible} and \textit{non-accessible} packer. The former enables the use of the actual code packer to generate packed samples, including all possible versions and configurations. This case is, therefore, ideal.
Hence, we consider the ``non-accessible packers'' scenario to be the general case.

\noindent \circled{2c} \textbf{Graphs Extraction.} \packhero extracts a \acrlong{cg} for each collected program. Our implementation relies on radare2~\cite{team_radare2_2023} to analyze and extract the \acrshort{cgs}. Each vertex of a \acrshort{cg} consists of 12 features extracted using radare2 (shown in Table~\ref{tab:features}). Furthermore, a heuristic designed to filter the unpacking stub part of the \acrshort{cg} is applied to simplify the topology of each graph (details in Subsection~\ref{subsec:heuristic_graphextraction}). \packhero collects the generated \acrshort{cgs} into a \acrfull{db} of graphs.

\begin{table}[b]
\centering
\small
\caption{Node features.}
\resizebox{0.9\textwidth}{!}{
\begin{tabular}{|l|p{12cm}|}
\hline
\textbf{Feature} & \textbf{Description} \\
\hline
\textit{type} &  Whether it is an internal function, a library imported function, or an entry point containing function. \\
\hline
\textit{size} & The size of the function in bytes. \\
\hline
\textit{real size} & The function size in bytes, including any padding. \\
\hline
\textit{is pure} & Indicates whether the function has any side effects such as modifying external variables or writing to files. \\
\hline
\textit{calling conventions} & The number of calling conventions used by the function. \\
\hline
\textit{number of basic blocks} & The number of basic blocks in the function. \\
\hline
\textit{number of instructions} & The total number of instructions in the function. \\
\hline
\textit{number of local variables} & The number of local variables declared within the function. \\
\hline
\textit{number of arguments} & The number of arguments of the function. \\
\hline
\textit{edges} & The number of edges between basic blocks. \\
\hline
\textit{indegree} & The in-degree of the function in the call graph. \\
\hline
\textit{outdegree} & The in-degree of the function in the call graph.  \\
\hline
\end{tabular}
}
\label{tab:features}
\vspace{-0.5cm}
\end{table}

\noindent \circled{3c} \textbf{Graph Matching Network Training.} \packhero identifies intrinsic similarities between extracted \acrshort{cgs} using a \acrfull{gmn}~\cite{li_graph_2019}, a specialized \acrfull{gnn}. The \acrshort{gmn} processes pairs of graphs and outputs a numeric vector (\textit{embedding}) for each graph. These embeddings result from information propagation between the two graphs, differing from traditional embedding techniques~\cite{hamilton_graph_2020} that compute embeddings solely from individual graphs.  To train the \acrshort{gmn}, we label graph pairs as ``similar'' if they originate from the same packer and ``dissimilar'' otherwise. The network is trained to minimize the distance between embeddings of similar graph pairs while maximizing the distance for dissimilar pairs.
The loss function is defined as $L(G_1, G_2) = \mathbb{E}_{(G_1, G_2,l)}\left[\max\{0, \gamma - l(1 - \text{cos}(G_1, G_2))\}\right]$, where $l\in \{-1, 1\}$ is the label associated with the pair of graphs $<G_1, G_2>$,
 $\gamma$ is a margin parameter and $cos$ is the cosine similarity~\cite{zaki_data_2020} between the two graphs. In this case, the $cos$ is intended as the cosine similarity between the two embeddings of size 256 extracted from the two graphs via the \acrshort{gmn}.
Lastly, $\mathbb{E}$ is the empirical risk we want to minimize, which can be done through stochastic gradient descent. The overall \acrshort{gmn} design follows the original implementation of the paper that introduced it~\cite{li_graph_2019}.
Finally, we propose a clustering approach to stabilize the number of matches required to identify each packer and improve the overall performance of the framework. Therefore, we also store the clusters and their respective medoids. Finally, we compute a cluster-specific threshold
$t_c = \frac{1}{n^2}\sum_{i=1}^{n}\sum_{j=1}^{n} cos(G_i, G_j) - \sigma$.
Namely, the average cosine distance between pairs of graphs belonging to the cluster minus the standard deviation. Such a threshold will then be used at inference time.

\subsection{Inference}\label{subsec:identification}

\begin{figure}[t]
  \centering
  \includegraphics[width=0.9\linewidth]{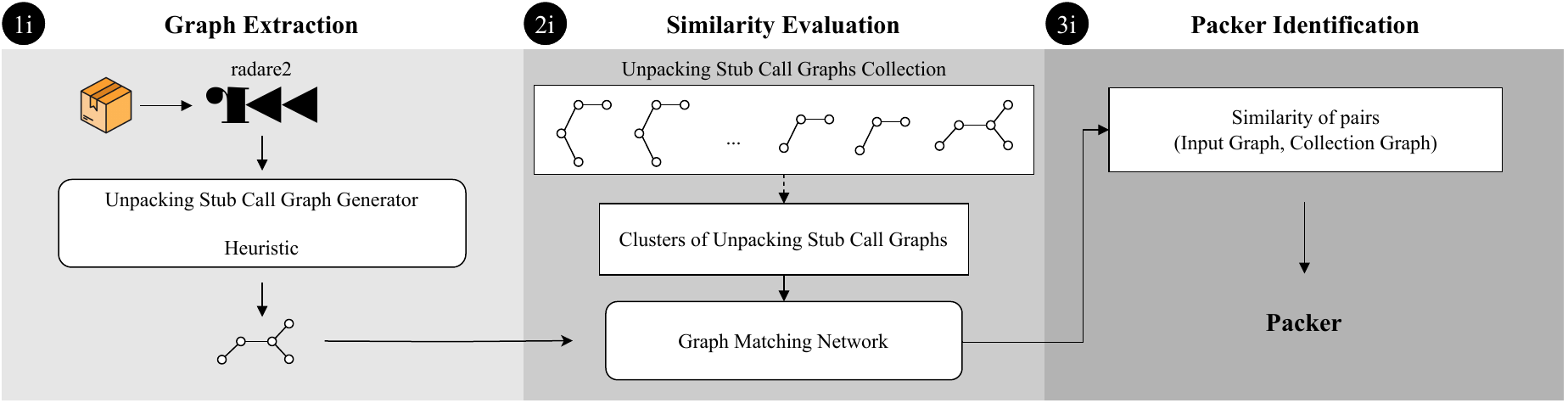}
  \caption{\packhero~Inference Workflow.}
  \label{fig:identification_workflow}
\vspace{-0.5cm}
\end{figure}

This phase comprises three steps, depicted in Fig.~\ref{fig:identification_workflow}.

\noindent \circled{1i} \textbf{Graph Extraction.} \packhero~must first obtain the \acrshort{cg} specific to the unpacking stub extracted through the previously mentioned heuristic. 

\noindent \circled{2i} \textbf{Similarity Evaluation.} The second step consists of evaluating the similarity between the embeddings computed by the \acrshort{gmn} for the input graph and the graphs in the \acrshort{db}. Comparing the input graph against all graphs in the \acrshort{db} may be computationally expensive and decrease the general identification performance. Hence, \packhero computes the cosine similarity between the input graph and the medoids associated with each computed cluster to select the ``closer'' clusters. In other words, each \packhero identification corresponds at least to $m$ \acrshort{gmn} inferences, where $m$ is the number of clusters. \packhero selects clusters represented by medoids with a positive cosine similarity with the input graph. Once the clusters are selected, \packhero evaluates the similarity between the input graph and each graph contained in the selected clusters.

\noindent \circled{3i} \textbf{Packer Identification.} 
Now, up to $m$ clusters are identified as potential matches, and the similarity between the input graph and all graphs within these $m$ clusters is computed. \packhero~identifies the packer with the highest score 
$s_p := \frac{\sum_{C \in \mathcal{C}_p} \sum_{G_c \in C} \mathbf{1}(\cos(G_{\text{input}}, G_c) \geq t_c)}{\max_{p \in \mathcal{P}} \sum_{C \in \mathcal{C}_{p}} \left| C \right|}$, where $G_{\text{input}}$ is the graph extracted from the input program, $C_p$ the set of selected clusters for a packer $p$, $G_c$ indicates a graph in cluster $C$, and $t_c$ the threshold for cluster $C$. $\mathbf{1}(\cos(G_{\text{input}}, G_c) > t_c)$ is a membership function that outputs 1 if the cosine similarity $\cos(G_{\text{input}}, G_c)$ is greater or equal to the threshold $t_c$, and 0 otherwise. Moreover, $\mathcal{P}$ is the set of all included packers in the selected clusters. Lastly, $\sum_{C \in \mathcal{C}_{p}} \left| C \right|$ is the cardinality of samples in the selected clusters from a packer $p$. If no cluster is sufficiently ``close'' to the input \acrshort{cg}, \packhero labels the packer as ``unknown''.

\subsection{Extracting the CG of the Unpacking Stub}\label{subsec:heuristic_graphextraction}

The \acrfull{cg} is a widely adopted structure~\cite{muchnick_advanced_1997}. It is also used in security-related tasks such as malware detection~\cite{segdroid}. Our approach is based on a principle of ``same packer, similar \acrshort{cgs}''~\cite{liu_2-spiff_2021}. To the best of our knowledge, we are the first to exploit this similarity directly. The adoption of \acrshort{cgs} offers several advantages. Their structure is straightforward to obtain~\cite{call_graphs}. Moreover, unlike other binary graph representations, a \acrshort{cg} represents the program at a higher level of abstraction. This makes it a compact yet information-rich program representation, which is particularly well-suited for a \acrshort{gmn}~\cite{li_graph_2019}.

\setlength{\textfloatsep}{0.3cm}
\begin{algorithm}[tb]
\scriptsize
\caption{Extract Unpacking Stub Call Graph}
\begin{algorithmic}[1]
\State $\mathcal{F}$: Set of all functions with call references, $\mathcal{E}$: Set of entry points in the binary
\Procedure{UnpackingStubCG}{$\mathcal{F}, \mathcal{E}$}
    \State $\mathcal{G} \gets \text{directed global call graph from $\mathcal{F}$}$, $\mathcal{C} \gets \emptyset$
    \For{$e \in \mathcal{E}$}
        \If{$\mathcal{G}.hasEdges(e)$}
            \State $\mathcal{C} \gets \mathcal{C} \cup \text{getConnectedComponent}(\mathcal{G}, e)$
        \EndIf
    \EndFor
    \State $\mathcal{G} \gets \mathcal{C} \neq \emptyset \ ? \ \mathcal{C} : \text{getComponent}(\mathcal{G}, \mathcal{E}) \cup \mathcal{E}$
    \If{$\mathcal{G}.isEmpty()$}
        \State $\mathcal{G} \gets \mathcal{E} \cup \text{externalLibraries}()$
    \EndIf
    \State \textbf{return} $\mathcal{G}$
\EndProcedure
\end{algorithmic}
\label{alg:heuristic}
\end{algorithm}

To better represent the logic behind a packer, it is necessary to filter the graph to get the unpacking stub. To systematically obtain this filtered \acrshort{cg}, we design a heuristic shown in Algorithm~\ref{alg:heuristic}. The intuitions behind it are: (i) the unpacking stub, or part of it (case of a multilayer packer), must be the first part of the code to be executed, and (ii) except for further obfuscation of the unpacking stub, a part of this routine is always statically visible. 
Therefore, the heuristic extracts the unpacking stub by exploiting the concept of connected components in undirected graphs, i.e., a subgraph where each pair of nodes is connected via a path~\cite{diestel_graph_2017}. 
Notice that \acrshort{cgs} are directed graphs, but the algorithm requires undirected ones, thereby we convert the \acrshort{cgs} into undirected graphs.
Given the packer could disrupt links between functions, it should create multiple connected components in the \acrshort{cg}. Thus, the idea is to extract the connected component containing the program entry point. At the same time, some packers affect the program entry point to make the analysis harder. For instance, analyzing \acrlong{cgs} extracted from binaries packed with ASPack~\cite{aspack}, we noticed the common part among all the graphs was a second connected component in addition to the single entry function node, which appears to be isolated. Thus, when the entry function is not connected to any other node, a second connected component is maintained in the graph along with the entry function. Otherwise, if the graphs have no edges (UPX~\cite{upx}), we keep only the program entry functions and any functions from external libraries. 
As the heuristic suggests, we do not consider a fixed number of functions for each graph. In our experimental evaluation, the average number of functions in the unpacking stubs is $\approx 3$.

\subsection{Graphs Clustering}\label{subsec:graphclustering}

\setlength{\textfloatsep}{0.3cm}
\begin{algorithm}[tb]
\scriptsize
\caption{Packer Call Graphs Clustering}
\label{alg:graphClustering}
\begin{algorithmic}[1]
\State DB: a collection of unpacking stub call graphs, \(\mathcal{M}\): \acrshort{gmn} trained model
\(\mathcal{P}\): Mapping from graphs to their respective packers
\Procedure{GetClustering}{$\text{DB},\mathcal{M},\mathcal{P}$}
    \State \(\mathcal{C} \gets \emptyset\) \Comment{Initialize clustering result}
    \For{each unique packer \(p \in \mathcal{P}\)}
        \State \(\text{DB}_p \gets \{ G \in \text{DB} \mid \mathcal{P}(G) = p \}\)
        \State \(\mathcal{D} \gets \text{initialize empty distance matrix for } \text{DB}_p\)
        \For{\(G_i, G_j \in \text{DB}_p, i \neq j\)}
            \State \(d \gets \text{cosineSimilarity}(\mathcal{M}(G_i, G_j))\)
            \State \(\mathcal{D}.update(d)\) \Comment{Update distance matrix with similarity}
        \EndFor
        \State \(\mathcal{C}_p \gets \text{hierarchicalClustering}(\mathcal{D})\)
        \For{\(C_j \in \mathcal{C}_p\)}
            \State \(C_j.\text{representative} \gets \text{medoid}(C_j)\)
        \EndFor
        \State \(\mathcal{C}.update(\mathcal{C}_p)\)
    \EndFor
    \State \textbf{return} \(\mathcal{C}\)
\EndProcedure
\end{algorithmic}
\end{algorithm}

Integrating a new packer requires collecting additional graphs. Without clustering, identifying a packer involves matching against \textit{all} graphs in the \acrshort{db}, increasing inference time as new packers are added. To ensure scalability, we introduce a clustering approach that reduces inference time and improves identification performance, as demonstrated in our Experimental Validation. Each cluster contains graphs from only a single packer, allowing the identification of potential sub-groups within the same packer. This \textit{packer unicity} is ensured by constructing the distance matrix in an intra-packer manner, as expressed in Algorithm~\ref{alg:graphClustering}. This approach can mitigate variations in unpacking stubs due to different configurations or versions of the same packer~\cite{li_packgenome_2023}. \packhero employs hierarchical clustering with a single linkage merge criterion, using a distance matrix derived from the trained \acrshort{gmn} as input. The silhouette score~\cite{zaki_data_2020} determines the optimal number of flat clusters for each packer. Finally, \packhero computes a medoid for each cluster, representing the graph with the minimal sum of dissimilarities to all other graphs in the cluster~\cite{zaki_data_2020}.

\section{Experimental Validation}\label{sec:exp_validation}

We evaluate \packhero through the following four research questions:

\mypar{RQ1} What is the minimum number of programs required for \packhero's configuration to recognize packers effectively? In addition, once configured, is \packhero able to recognize different packers?

\mypar{RQ2} How does the clustering-based approach impact \packhero's performance and scalability compared to its non-clustering version?

\mypar{RQ3} Given an already configured \packhero, how many samples does it require to successfully integrate a new, unseen packer, and how does this integration compare to other \acrshort{ml}-based tools?

\mypar{RQ4} Does \packhero perform better than signature-based tools?

\subsection{Experimental Setup}
\label{subsec:experimental_setup}

\mypar{Dataset} We use the \textit{lab} dataset from H. Aghakhani et al.\cite{aghakhani_when_2020}, created by repacking \acrfull{pes} from benign and malicious Windows x86 software collected from a commercial anti-malware vendor and the EMBER dataset\cite{anderson2018ember}. The samples were repacked using nine widely recognized packers: kkrunchy, MPRESS, Obsidium, PECompact, PELock, Petite, tElock, Themida, and UPX.
The dataset includes different packer families. Following a \acrshort{sota} taxonomy~\cite{ugarte-pedrero_sok_2015}, it covers Type-I (e.g., UPX), Type-III (e.g., PECompact), and \acrshort{vm}-based packers (e.g., Themida). To replicate Experiment II from the original work, we apply the same undersampling strategy, resulting in 15,353 samples per packer. We randomly select 10\% of the undersampled dataset while preserving the distribution of malware, benign programs, and packers. This subset, referred to as the \textit{lab-10} dataset, excludes 10 outliers (\acrshort{cgs} with more than 500 nodes). In Section~\ref{subsec:train_set_size}, we test \packhero with the \textit{RGD} dataset from PackGenome~\cite{li_packgenome_2023}, which consists of three manually constructed programs, compiled from 2-5 lines of C code and packed with several versions and configurations of 20 off-the-shelf packers. To assess transferability, we select only the \textit{RGD} packers also present in \textit{lab-10}.

\mypar{Hyperparameter Tuning} We optimize the hyperparameters of the \acrshort{gmn} by maximizing intra-packet similarity using a grid search approach.

\mypar{Evaluation Metrics} We evaluate \packhero using metrics~\cite{zaki_data_2020} such as precision, recall, F1-score, accuracy, \acrfull{fpr}, and the unknown rate, which indicates how often \packhero fails to recognize a packer. We also measure the \textit{Average Number of Inference Calls}, representing the average calls \packhero makes to the \acrshort{gmn} during identification.

\begin{figure}[t]
  \centering
  \begin{subfigure}{0.95\linewidth}
    \centering
    \includegraphics[width=\linewidth]{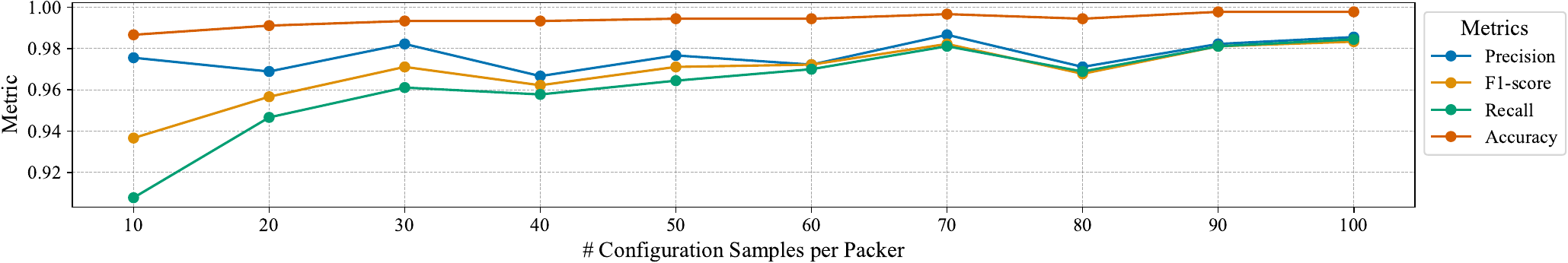}
  \end{subfigure}
  
  \begin{subfigure}{0.95\linewidth}
    \centering
    \includegraphics[width=\linewidth]{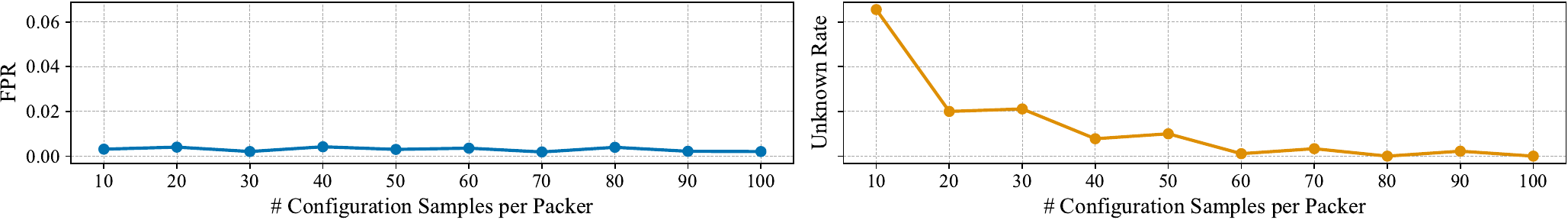}
  \end{subfigure}
  
  \caption{Metrics trends varying the number of programs to configure \packhero.}
  \label{fig:trainingsize_combined}
\end{figure}

\subsection{Effective Identification (RQ1)}
\label{subsec:train_set_size}

\mypar{Configuration} \packhero requires a configuration phase starting with the step of collecting programs. If the packer is accessible, i.e., we can use it to craft new samples, the configuration becomes trivial and we can obtain as many programs as we need. However, if the packer is not accessible, we need to collect packed programs in the wild. These programs can be malware or benign, and it is uncertain whether we can find them in large quantities. Therefore, we test \packhero in a scenario with a limited number of programs, defining the number of programs needed for each packer to achieve a good average performance.
From the \textit{lab-10}, we select 100 programs for each packer to configure \packhero, maintaining the original distribution of malware and benign programs. We use the remaining programs for testing. Starting from the 100 programs for each packer, we gradually eliminate 10 programs and create 10 different collections of gradually smaller sizes. We use these 10 collections to configure \packhero. Then, we test our approach, configured with different ``training'' sizes, using the same test dataset. The metrics we use to evaluate \packhero~in this experiment are precision, recall, F1-score, accuracy, \acrshort{fpr}, and the unknown rate. Each plot in Fig.~\ref{fig:trainingsize_combined} shows the macro-average results, i.e., the average results among the 9 packers in the dataset. Looking at the precision, F1-score, recall, and accuracy, the tool does not perform badly even with only 10 programs per packer. Furthermore, starting from 30 samples per packer, \packhero maintains all metrics above 0.96. In addition, the plot shows precision and recall converge in the long run. We also notice that from the configuration of 70 samples for packers, \packhero achieves a good balance between precision and recall, which means a good tradeoff between \acrfull{fps} and \acrfull{fns}. As regards the \acrshort{fpr} and the unknown rate trends, they follow all the other metrics. The \acrshort{fpr} is overall low and always below 0.00418, i.e., 0.41\% of \acrshort{fps}. We also observe higher unknown rates for lower ``training sizes'', which means \packhero becomes more confident in his choices as the ``training'' size increases. The plot fluctuations are because the experiment was done with a single run due to the computational cost of training and testing with 10 different configurations. 
However, a single run places \packhero in a realistic scenario with limited samples and no ability to select the most suitable ones for tool configuration.

\begin{table}[b]
\centering
\caption{\packhero~performance on lab-10. UR denotes the Unknown Rate.}
\resizebox{0.55\textwidth}{!}{
\begin{tabular}{lcccccccc}
\toprule
\textbf{Packer} & \textbf{\#Samples} & \textbf{UR} & \textbf{FPR} & \textbf{Prec} & \textbf{Rec} & \textbf{F1} & \textbf{Acc}\\
\hline
kkrunchy & 1435 & 0.0 & 0.0001 & 1.00 & 0.99 & 1.00 & 1.00\\
MPRESS & 1435 & 0.0 & 0.0018 & 0.99 & 0.98 & 0.98 & 1.00\\
Obsidium & 1435 & 0.0 & 0.0031 & 0.98 & 0.98 & 0.98 & 0.99\\
PECompact & 1434 & 0.0 & 0.0012 & 0.99 & 1.00 & 0.99 & 1.00\\
PELock & 1435 & 0.0 & 0.0002 & 1.00 & 0.96 & 0.98 & 1.00\\
Petite & 1434 & 0.0 & 0.0001 & 1.00 & 0.99 & 0.99 & 1.00\\
tElock & 1432 & 0.0 & 0.0107 & 0.92 & 0.98 & 0.95 & 0.99\\
Themida & 1433 & 0.0 & 0.0007 & 0.99 & 1.00 & 0.99 & 1.00\\
UPX & 1432 & 0.0 & 0.0006 & 1.00 & 0.98 & 0.99 & 1.00\\
\bottomrule
\textbf{macro-avg} & - & \textbf{0.0} & \textbf{0.0021} & \textbf{0.986} & \textbf{0.984} & \textbf{0.983} & \textbf{0.998}\\
\bottomrule
\end{tabular}
}
\label{tab:packerszoom}
\end{table}

\mypar{Effectiveness on different packers} Here, we zoom in on the results obtained from the best configuration in the previous experiment, namely the one with 100 samples for packers. The test set is the remaining part of the dataset. Table~\ref{tab:packerszoom} shows that \packhero~performs very well on each packer in the dataset. We have a near-maximum accuracy in general and equally good results in all other metrics for other packers. The only exception is tElock, which is found to have a higher \acrshort{fpr} than the others. This result has chain effects on precision and F1-score. An answer can be found by looking at the clusters' composition.
In particular, tElock produces two clusters of 1 and 99 samples. In its current version, \packhero merges single-sample clusters with the nearest one. As a result, for tElock, we obtain a single cluster consisting mostly of similar samples, along with one slightly different sample, which lowers intra-cluster similarity. This leads to a reduced threshold ($\approx 0.10$ lower than others) and lower confidence in the choice. This observation suggests that treating single-element clusters as outliers and excluding them could enhance \packhero's performance.

\begin{table}[b]
    \centering
    \caption{\packhero performance on RGD. A configuration is considered identified if all samples from that configuration are correctly classified.}
    \resizebox{0.78\textwidth}{!}{
    \begin{tabular}{lc}
        \toprule
        \textbf{Packer}  & \textbf{Versions (\#Identified Configurations/\#Configurations)} \\
        \hline
        kkrunchy    & v0.23alpha \textbf{(0/2)}, v0.23alpha2 \textbf{(1/1) }   \\
        MPRESS      & v1.27 \textbf{(1/1)}, v2.18 \textbf{(1/1)}, v2.19 \textbf{(1/1)} \\
        Obsidium    & v1.5 \textbf{(7/7)}                     \\
        PEcompact   & v3.02.2 \textbf{(18/19)}, v3.11 \textbf{(10/12)}   \\
        PElock      & v1.06 \textbf{(0/5)}                    \\
        Petite      & v2.4 \textbf{(5/5)}                     \\
        Themida     & v2.37 \textbf{(8/9)}, v2.39 (\textit{Winlicense}) \textbf{(8/9)} v3.04 \textbf{(0/5)}      \\
        UPX         & v1.00 \textbf{(4/4)}, v1.20 \textbf{(8/8)}, v1.25 \textbf{(4/4)}, v2.00 \textbf{(5/5)}, v3.09 \textbf{(8/8)}, v3.96 \textbf{(8/8)} \\
        \bottomrule
    \end{tabular}
    }
    \label{tab:packer_versions}
\end{table}

\mypar{Different versions and configurations} The \textit{lab-10} dataset includes only one configuration and version per packer. To evaluate transferability, we test \packhero's best configuration (\textit{lab-10}) on the \textit{RGD} dataset~\cite{li_packgenome_2023}, which contains multiple versions and configurations for each packer—except for tElock, which is not included in the PackGenome evaluation. Table~\ref{tab:packer_versions} presents the configurations identified by \packhero for each version in \textit{RGD}. An ``identified configuration'' occurs when \packhero recognizes all samples, while non-identified configurations show a 0\% identification rate, likely due to differences in the unpacking stub. Overall, \packhero generalizes across 16 out of 19 different versions, despite being configured with only a single version and configuration per packer.

\mypar{Answer to RQ1} The number of packed programs required to configure \packhero depends on the desired performance level. With just 10 samples per packer, \packhero achieves a minimum macro-average F1 score of 93.7\% and accuracy of 98.7\%. Increasing the sample size to 30 can further improve recall and F1-score while maintaining high precision and accuracy.
Table~\ref{tab:packerszoom} demonstrates \packhero's ability to effectively identify multiple packers from different families. Given the dataset's composition, \packhero successfully integrates and recognizes packers of varying complexity in both packed malware and benign programs. Specifically, based on the taxonomy by Ugarte-Pedrero et al.~\cite{ugarte-pedrero_sok_2015}, \packhero performs well on Type-I (UPX), \acrshort{vm}-based (Themida), and Type-III (PECompact) packers.

\subsection{Clustering Effectiveness (RQ2)}
\label{subsec:clustering_effectiveness}

\begin{figure}[t]
  \centering
  \begin{subfigure}{0.95\linewidth}
    \centering
    \includegraphics[width=\linewidth]{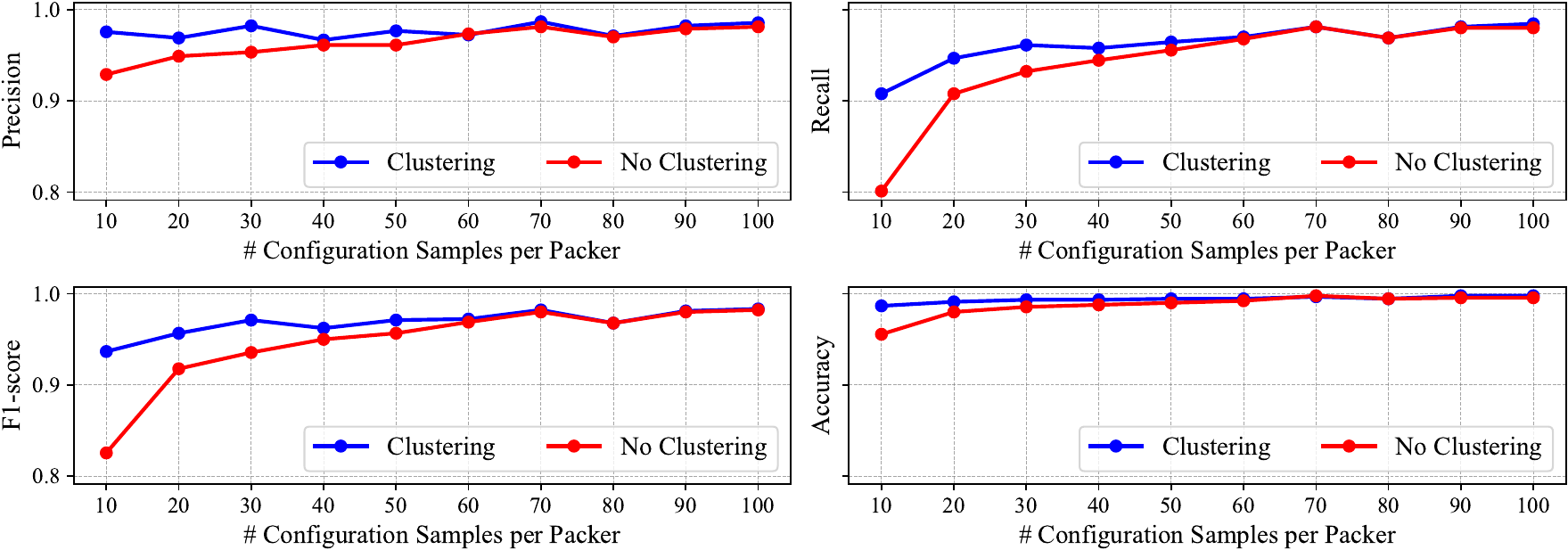}
  \end{subfigure}
  
  \begin{subfigure}{0.95\linewidth}
    \centering
    \includegraphics[width=\linewidth]{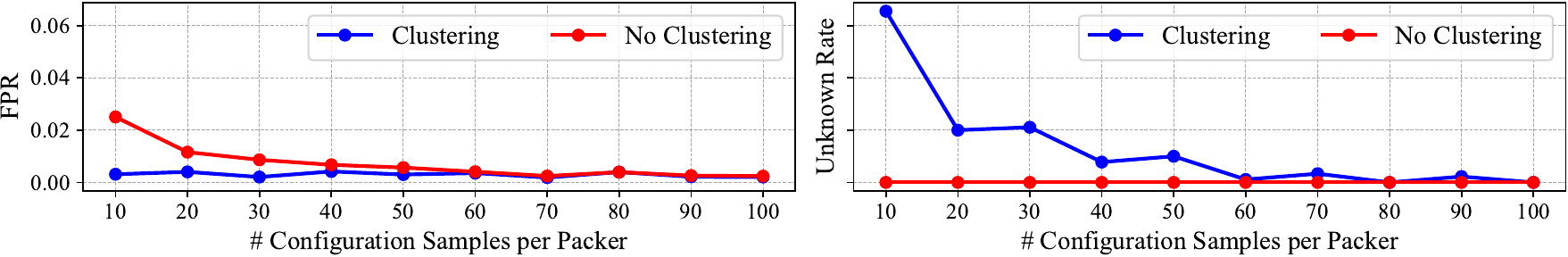}
  \end{subfigure}
  \caption{Comparison between \packhero with and without clustering.}
  \label{fig:clusteringeffectiveness_combined}
\end{figure}

We evaluate the impact of the clustering approach on \packhero's performance and scalability. To do this, we replicate the experiment from Section~\ref{subsec:train_set_size} without the clustering layer: \packhero evaluates similarity with all graphs in the \acrshort{db} and computes packer-specific thresholds instead of cluster-specific ones.  
Fig.~\ref{fig:clusteringeffectiveness_combined} illustrates the performance gap between \packhero with and without clustering. This gap is more pronounced for smaller training set sizes and narrows as the training set size increases. Without clustering, the unknown rate consistently drops to 0, as the \textit{unknown} classification (explained in Section~\ref{sec:packhero}) depends on the similarity step involving clusters' medoids. However, the absence of clustering increases \acrlong{fps}. These results demonstrate that incorporating clustering into \packhero's workflow is effective and that the chosen medoids are good representatives of subgroups within the same packer.

\begin{table}[b]
\centering
\caption{Average number of inference calls made by \packhero to the \acrfull{gmn} during packer identification of a single sample.}
\resizebox{0.61\textwidth}{!}{
\begin{tabular}{|c|c|c|c|c|}
\hline
\textbf{Configuration} & \textbf{\#Clusters} & \multicolumn{2}{c|}{\textbf{Clustering}} & \textbf{No Clustering} \\
\cline{3-4}
& & \textbf{Ideal Values} & \textbf{Real Values} & \\
\hline
10 & 14 & 24.00 & 24.14 $\pm$ 1.33 & 90.00 $\pm$ 0.00 \\
20 & 13 & 33.00 & 33.02 $\pm$ 0.17 & 180.00 $\pm$ 0.00 \\
30 & 17 & 47.00 & 47.03 $\pm$ 1.19 & 270.00 $\pm$ 0.00 \\
40 & 16 & 56.00 & 60.59 $\pm$ 4.68 & 360.00 $\pm$ 0.00 \\
50 & 13 & 63.00 & 64.31 $\pm$ 2.65 & 450.00 $\pm$ 0.00 \\
60 & 14 & 74.00 & 82.68 $\pm$ 7.48 & 540.00 $\pm$ 0.00 \\
70 & 11 & 81.00 & 83.71 $\pm$ 3.84 & 630.00 $\pm$ 0.00 \\
80 & 16 & 96.00 & 107.54 $\pm$ 13.35 & 720.00 $\pm$ 0.00 \\
90 & 13 & 103.00 & 107.30 $\pm$ 3.57 & 810.00 $\pm$ 0.00 \\
100 & 17 & 117.00 & 124.47 $\pm$ 7.78 & 900.00 $\pm$ 0.00 \\
\hline
\end{tabular}
}
\label{tab:avg_inferencecalls}
\end{table}

Table~\ref{tab:avg_inferencecalls} shows the average number of inference calls to the \acrshort{gmn} during the identification, i.e., the average number of matched graphs needed to identify a packer from a program. Removing the clustering approach, the similarity with the input graph is evaluated with the entire collection. Thus, the metric is always equal to the size of the entire collection. In contrast, the values we empirically obtain with the use of the clustering approach show are close to the ideal number of inference calls to identify the packer.
The ideal number of inference calls in the presence of the clustering approach is represented by the sum of $m$ (number of clusters) and the number of programs per packer stored in the \acrshort{db}. Thus, we can state that, involving the clustering approach, the \packhero's number of inference calls does not depend on the number of packers but only on the number of samples for each packer stored in the \acrshort{db}. To further emphasize the significance of the results shown in Table~\ref{tab:avg_inferencecalls}, it is important to consider the inference time for our \acrshort{gmn}. Indeed, while this network's expressive power surpasses that of its alternatives, it comes with the trade-off of increased temporal complexity. In this experiment, the average single inference time recorded was $1.76ms$. The time difference observed in the tests conducted on 9 packers is not substantial between the version with and without clustering. However, in a scenario with $200$ packers and $100$ samples in the \acrshort{db} for each packer, the time required for identification would be $\approx 35s$ without clustering but only $21ms$ with it.

\mypar{Answer to RQ2} The clustering approach improves \packhero's performance, especially with limited samples per packer, while its effectiveness remains unaffected by the number of recognized packers, ensuring scalability.

\subsection{Integration of New Packers (RQ3)}
\label{subsec:integration_new_packers}

As discussed in Subsection~\ref{subsec:train_set_size}, to demonstrate the integration is always feasible we have to deal with a scenario in which we need to collect programs in the wild to integrate the new packer. Thus, we have to face the possibility that these programs are not numerous.
We have already tested \packhero in a scenario with few samples for its configuration. Here, we aim to evaluate how many samples \packhero needs to ``integrate'' a new packer with the entire system already configured. The process of integration corresponds to the ``configuration'' of \packhero described in Subsection~\ref{fig:config_workflow}. Since the workflow includes a \acrfull{gmn}, we can avoid re-training the model from scratch. Indeed, given that we use a \acrfull{nn}, it can be updated through fine-tuning, i.e., partially re-training on new samples. At the same time, \packhero may even allow us to avoid fine-tuning the model altogether. Specifically, a new packer can be integrated into \packhero without re-training the \acrshort{gmn}, simply by adding its corresponding graphs to the \acrshort{db}. However, this approach is feasible only if the collected graphs for the packer are sufficiently homogeneous. Currently, we assume that manual intervention was previously performed on the packer samples to be integrated, which we assume are always correctly labeled.

\begin{figure}[t]
  \centering
  \includegraphics[width=0.9\linewidth]{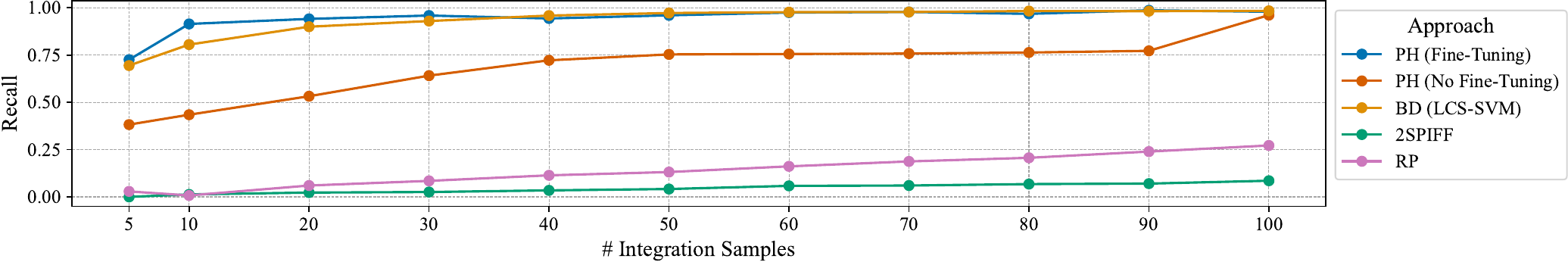}
  \caption{Recall trend increasing the number of integration samples; \packhero~(Fine-Tuned and Not) vs \acrshort{ml}-based Tools.}
  \label{fig:againstml_performance}
\end{figure}

We evaluate \packhero with and without fine-tuning the \acrshort{gmn}. We train a \acrshort{gmn} for each packer, excluding it from the training set, which consists of 100 samples per remaining packer. Then, we integrate samples from the ``unseen'' packer. In the version without fine-tuning, we add the new packer’s graphs directly to the \acrshort{db}. In the fine-tuned version, we fine-tune the \acrshort{gmn} using the new graphs before integration.
To compare \packhero with \acrshort{sota} tools, we replicate the experiment using three \acrshort{ml}-based approaches from Table~\ref{tab:sota}: Randomness Profiles~\cite{sun_pattern_2010}, Binary Diffing~\cite{binary_diffing} (best-performing version: LCS-SVM), and 2SPIFF~\cite{liu_2-spiff_2021}. As implementations of these tools are unavailable, we reimplement them to the best of our ability and validate the implementations by comparing the achieved accuracy on the remaining packers in this experiment.
For a fair evaluation, we train them using a stratified 80-20 split of the \textit{lab-10} dataset. Therefore, for each \acrshort{ml}-based tool, we train each model by excluding one packer, using 80\% of the dataset. This subset includes the 100 samples per packer used to train \packhero.
However, none of these approaches utilize a \acrshort{nn}. Therefore, we are required to entirely re-train their models each time we want to integrate a new packer. Ultimately, both our method and the other approaches are tested using the test set from the 80-20 split, which corresponds to the stratified 20\% of \textit{lab-10}, equal to 307 samples for each packer. Results specific to each packer are obtained by testing the samples from that particular packer on the updated tools. By evaluating separately for each packer, the metric we use is recall, which is equivalent to accuracy in this experimental setup.
In Fig.~\ref{fig:againstml_performance}, we show the average recall trends for \packhero with and without fine-tuning and compare them against all other \acrshort{ml}-based packer identifiers. As the figure shows, 2SPIFF and Randomness Profiles (RP) perform very badly, even in the best configuration. In contrast, both the fine-tuned \packhero and Binary Diffing (BD) with LCS-SVM achieve very good performance and consistency using a small number of samples. Starting from the 40 samples, their performance is aligned, except for small fluctuations due to the single run. However, as the plot shows, \packhero reaches a high recall before BD. Indeed, using 5, 10, 20, and 30 samples to integrate the new packer, \packhero performs better. Similarly, \packhero without fine-tuning also exhibits good performance, although not as good as the other two methods. However, the average performance hides the results for single-packers.
Indeed, removing two packers out of nine (Obsidium and Petite) the non-fine-tuned \packhero achieves performance very close to the fine-tuned version and BD starting from 50 integration samples. 
This result shows that fine-tuning the \acrshort{gmn} is unnecessary for low-heterogeneity graphs, saving computation.

\begin{figure}[t]
  \centering
  \includegraphics[width=0.9\linewidth]{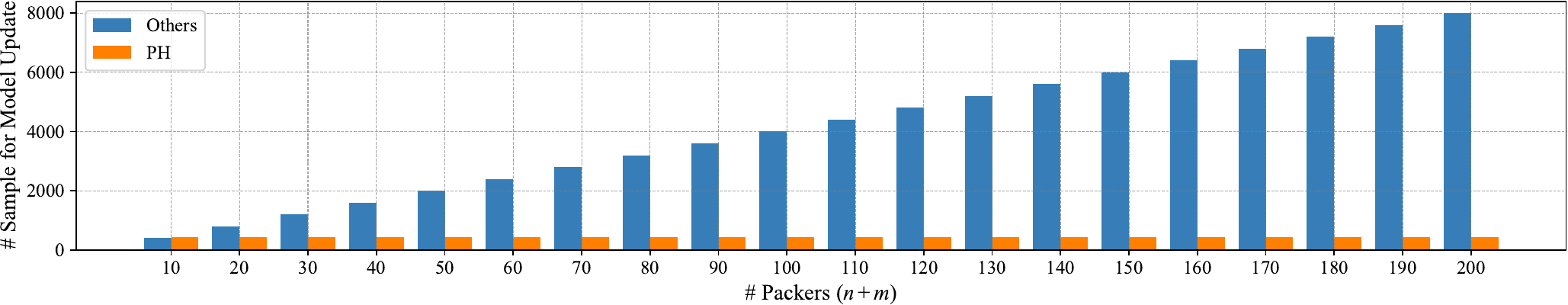}
  \caption{\# Samples involved in the integration ($m = 10$ new packers, $l = 40$ samples per packer, $n$ already integrated packers).}
  \label{fig:againstml_trainingcost}
\end{figure}

Additionally, all \acrshort{ml}-based approaches require complete retraining to integrate new packers, a scalability issue \packhero avoids. To demonstrate this, we simulate packer integration during the tool lifecycle. Assuming integration cost depends on the number of samples used, we define the cost function $f(n,m,l)$, where $n$ is the number of known packers, $m$ the new packers, and $l$ the samples per packer.  
For other tools, $f(n,m,l) = (n + m) \cdot l$, while \packhero’s cost is $f(n,m,l) = m \cdot l$, as it only depends on new packers. Fig.~\ref{fig:againstml_trainingcost} simulates $m = 10$ and $l = 40$, showing \packhero’s constant integration cost, unlike other tools, where cost grows linearly with the number of recognized packers.
This result demonstrates that \packhero scales effectively in realistic scenarios where new packers must be integrated over time.

\mypar{Answer to RQ3} \packhero achieves strong results in integrating a new ``unseen'' packer with as few as 10 samples. Among 9 tested packers, \packhero outperforms the three other selected \acrshort{ml}-based tools for integration sample counts less than 30. Furthermore, a simulation comparing retraining and fine-tuning shows that \packhero’s integration cost function remains constant as new packers are added, unlike other \acrshort{ml}-based approaches, whose integration costs grow linearly with the number of recognized packers.

\subsection{Signature-Based Tool Comparison (RQ4)}
\label{subsec:signaturebased_comparison}

\begin{table}[b]
\centering
\caption{\packhero~vs \acrshort{die} vs PackGenome on lab-10.}
\resizebox{0.55\textwidth}{!}{
\begin{tabular}{cccccccc}
\hline
\multicolumn{2}{c}{} & \multicolumn{3}{c}{\textbf{Recall}} & \multicolumn{3}{c}{\textbf{FPR}} \\
\cmidrule(lr){3-5} \cmidrule(lr){6-8}
\rowcolor{white}
 & \textbf{\#Samples} & \textbf{PH} & \acrshort{die} & PG & \textbf{PH} & \acrshort{die} & PG\\
\hline
kkrunchy & 1435 & 0.99 & 1.0 & 1.0 & 0.0001 & 0.0 & 0.0\\
MPRESS & 1435 & 0.98 & 1.0 & 1.0 & 0.0018 & 0.0008 & 0.0034\\
Obsidium & 1435 & 0.98 & 1.0 & 1.0 & 0.0031 & 0.0002 & 0.0000\\
PECompact & 1434 & 1.0 & 1.0 & 1.0 & 0.0012 & 0.0 & 0.0005\\
Petite & 1434 & 0.99 & 0.99 & 0.99 & 0.0001 & 0.0 & 0.0\\
Themida & 1433 & 1.0 & 0.92 & 0.31 & 0.0007 & 0.0 & 0.0499\\
UPX & 1432 & 0.98 & 1.0 & 1.0 & 0.0006 & 0.0002 & 0.0011\\
\hline        
\end{tabular}
}
\label{tab:signaturecomparison}
\end{table}

In this experiment, we compare the performance of \packhero with \acrlong{sota} signature-based methods (\acrfull{die} and PackGenome).
\acrshort{die}~\cite{detectiteasy} is currently the best-performing and most signature-rich packer identifier. Here, we use its latest available version (v3.10). PackGenome~\cite{li_packgenome_2023} is the best tool for automating the extraction of signatures for packer detection, generating YARA~\cite{yara} rules that can be later used to identify packers. The authors of PackGenome have already compared their framework against \acrshort{die}, but we include the results for both approaches for completeness.
To compare the three frameworks, we use the \acrfull{fpr} and recall. We focus on recall because it is the most representative metric when comparing tools designed not just for identification but also for detection (as for \acrshort{die} and PackGenome). We extract all results from the \textit{lab-10} dataset, removing only the 100 samples for each packer used to configure \packhero. We test \acrshort{die} using its command-line version, while we test PackGenome by loading the \yara rules provided in the original work. Since \acrshort{die} does not include PELock signatures and PackGenome's authors did not perform experiments on PELock and tElock, we discard these two packers for this comparison.
In Table~\ref{tab:signaturecomparison}, we show the comparison results. Starting with the recall, the three tools perform very similarly, although the two signature-based tools generally exhibit the highest recall. Attention must particularly be directed towards Themida, which poses significant challenges for both \acrshort{die} and PackGenome, as depicted in the table. An interesting observation is that the matched signatures from \acrshort{die} are related to the same version of Winlicense/Themida, specifically Themida with Trial/Licensing options~\cite{winlicense}. Despite PackGenome including signatures for the same version of this packer, it shows a recall of 0.31. This result motivates the entire work. Indeed, Themida/Winlicense employs advanced (dynamic) evasive behaviors in its packed programs. Furthermore, it is the \acrshort{vm}-based packer used during our evaluation. As explained in Section~\ref{subsec:signature_background}, PackGenome extracts \yara rules by tracing instructions during their execution. Consequently, it is likely to struggle with the evasive behaviors introduced by Themida into the binary during the packing process. Additionally, PackGenome appears to face challenges due to the inherent nature of this packer. Indeed, the result suggests that both the signature itself and its automatic extraction encounter difficulties when dealing with this packer family. Finally, looking at \acrshort{fpr}, PackGenome confirms its issues with Themida/Winlicense but shows in-line results for the other packers. \packhero~demonstrates a low \acrshort{fpr} on average, while \acrshort{die} generates the fewest \acrshort{fpr}.

\mypar{Answer to RQ4} \packhero matches \acrshort{sota} signature-based tools in accuracy and significantly outperforms them on \acrshort{vm}-based packers with advanced evasive behaviors like Themida/Winlicense, demonstrating our approach's effectiveness.

\section{Limitations and Future Work} \label{sec:lim_futwork}

\packhero relies on heuristics to extract filtered \acrshort{cgs}. Unpacking stubs play a crucial role in the analysis but other code segments might also contribute to the \acrshort{cg}'s structure. Even if this work demonstrates that a few statically visible functions are often sufficient to determine the packer's identity, the exact number of functions considered for each \acrshort{cg} remains an open aspect. \packhero directly exploits disassembly and function identification, which are inherently challenging problems, especially in the context of malware due to obfuscation techniques, indirect branch resolution, and evasive behaviors. Furthermore, \packhero is potentially vulnerable to adversarial attacks that manipulate the \acrshort{cg} to evade identification. An adversary could, for instance, obfuscate the \acrshort{cg} by inserting bogus functions, modifying calls, or hiding call targets, significantly complicating packer identification. Additionally, different dynamic evasive behaviors implemented by malware could further impact the accuracy of the extracted \acrshort{cg}. 
Hence, future work may study heuristics to resist adversarial attacks by evaluating \acrshort{cg} obfuscation to identify which aspects of our heuristics and features are most susceptible to evasion. In our study, we selected radare2 as the disassembler due to its ease of use and integration. However, recent research has demonstrated that several other open-source disassemblers outperform radare2 performance~\cite{sokdisassemblers}. This reliance, while currently effective, necessitates further investigation, particularly in scenarios involving adversarial manipulation of the unpacking stub or \acrshort{cg} structure. Finally, \packhero currently focuses on \textit{packer identification} but does not determine whether a sample is packed (\textit{detection}). Preliminary analyses revealed a notable number of \acrlong{fps} when analyzing non-packed samples, indicating the need for further improvements in this area. Therefore, future work will also address the \textit{packer detection} problem.
\section{Conclusion}\label{sec:conclusions}

This paper introduced \packhero, a packer identifier that leverages a heuristic to extract a \acrfull{cg} representing the unpacking routine of a program. Using a clustering approach to enhance performance and reduce the search space, \packhero evaluates the similarity between the extracted \acrshort{cg} and labeled \acrshort{cgs} stored in a \acrshort{db}, employing a \acrfull{gmn} to compute these similarities and identify the packer.
Evaluated on a public dataset of packed benign and malicious programs re-packed multiple times, \packhero meets all key requirements for a novel packer identifier: high accuracy, efficient packer integration, evasive behavior management, and scalability. Relying exclusively on static analysis, \packhero integrates new packers effectively, achieving strong performance with as few as 10 samples, while eliminating the limitations of dynamic analysis, particularly against dynamic evasive behaviors. For some packers, it avoids fine-tuning the \acrshort{gmn}, and when fine-tuning is needed, it converges faster than other ML-based tools. Its integration cost remains constant throughout its lifecycle, unlike other methods, where costs grow linearly with the number of packers recognized.
\packhero performs comparably to signature-based tools, the current best-performing solutions for packer identification, and significantly outperforms \acrshort{sota} approaches on Themida, a \acrshort{vm}-based packer employing advanced dynamic evasive behaviors.

{\small \mypar{Acknowledgements} This work was partially supported by Project FARE (PNRR M4.C2.1.1 PRIN 2022, Cod. 202225BZJC, CUP D53D23008380006, Avviso D.D 104 02.02.2022) and Project SETA (PNRR M4.C2.1.1 PRIN 2022, Cod. P202233M9Z, CUP F53D23009120001, Avviso D.D 1409 14.09.2022) under the Italian NRRP MUR program, and by Project SERICS (PE00000014) under the MUR National Recovery and Resilience Plan, all funded by the European Union - NextGenerationEU.}

\bibliographystyle{splncs04}
\bibliography{bibliography}

\begin{thebibliography}{10}
\providecommand{\url}[1]{\texttt{#1}}
\providecommand{\urlprefix}{URL }
\providecommand{\doi}[1]{https://doi.org/#1}

\bibitem{aghakhani_when_2020}
Aghakhani, H., Gritti, F., Mecca, F., Lindorfer, M., Ortolani, S., Balzarotti, D., Vigna, G., Kruegel, C.: When {Malware} is {Packin}' {Heat}; {Limits} of {Machine} {Learning} {Classifiers} {Based} on {Static} {Analysis} {Features}. In: Proceedings of {Symposium} on {Network} and {Distributed} {System} {Security} ({NDSS}) (Feb 2020)

\bibitem{entropy_2}
Al-Anezi, D.M.M.K.: Generic packing detection using several complexity analysis for accurate malware detection. International Journal of Advanced Computer Science and Applications  \textbf{5}(1) (2014). \doi{10.14569/IJACSA.2014.050102}

\bibitem{yara}
Alvarez, V.M.: Yara. \url{https://virustotal.github.io/yara/} (2024), accessed: 2024-04-15

\bibitem{anderson2018ember}
Anderson, H.S., Roth, P.: Ember: an open dataset for training static pe malware machine learning models. arXiv preprint arXiv:1804.04637  (2018)

\bibitem{aspack}
{ASPack Software}: {A}{S}{P}ack {S}oftware - {A}pplication for compression, packing and protection of software. \url{http://www.aspack.com/} (2024), accessed: 2024-04-15

\bibitem{call_graphs}
Callahan, D., Carle, A., Hall, M., Kennedy, K.: Constructing the procedure call multigraph. IEEE Transactions on Software Engineering  \textbf{16}(4),  483--487 (1990). \doi{10.1109/32.54302}

\bibitem{diestel_graph_2017}
Diestel, R.: Graph {Theory}. Springer, 5th edn. (2017)

\bibitem{sliding_windows}
Ebringer, T., Sun, L., Boztas, S.: A fast randomness test that preserves local detail. In: Proceedings of the 18th Virus Bulletin International Conference. pp. 34--42. Virus Bulletin Ltd (2008)

\bibitem{survey_anti_analysis}
Egele, M., Scholte, T., Kirda, E., Kruegel, C.: A survey on automated dynamic malware-analysis techniques and tools. ACM Comput. Surv.  \textbf{44}(2) (mar 2008). \doi{10.1145/2089125.2089126}

\bibitem{galloro_systematical_2022}
Galloro, N., Polino, M., Carminati, M., Continella, A., Zanero, S.: A {Systematical} and longitudinal study of evasive behaviors in windows malware. Computers \& Security  \textbf{113},  102550 (Feb 2022). \doi{10.1016/j.cose.2021.102550}

\bibitem{hamilton_graph_2020}
Hamilton, W.L.: Graph {Representation} {Learning}. Synthesis Lectures on Artificial Intelligence and Machine Learning  \textbf{14}(3),  1--159 (2020), publisher: Morgan and Claypool

\bibitem{entropy_1}
Hamrock, J., Lyda, R.: Using entropy analysis to find encrypted and packed malware. IEEE Security \& Privacy  \textbf{5}(02),  40--45 (mar 2007). \doi{10.1109/MSP.2007.48}

\bibitem{detectiteasy}
Horsicq: Detect it easy. \url{https://github.com/horsicq/Detect-It-Easy} (2024), accessed: 2024-04-15

\bibitem{packed_similarity}
Jacob, G., Comparetti, P., Neugschwandtner, M., Kruegel, C., Vigna, G.: A static, packer-agnostic filter to detect similar malware samples. In: International Conference on Detection of intrusions and malware, and vulnerability assessment. vol.~7591 (01 2010). \doi{10.1007/978-3-642-37300-8_6}

\bibitem{binary_diffing}
Kim, Y., Paik, J.Y., Choi, S., Cho, E.S.: Efficient svm based packer identification with binary diffing measures. In: 2019 IEEE 43rd Annual Computer Software and Applications Conference (COMPSAC). vol.~1, pp. 795--800 (2019). \doi{10.1109/COMPSAC.2019.00117}

\bibitem{li_packgenome_2023}
Li, S., Ming, J., Qiu, P., Chen, Q., Liu, L., Bao, H., Wang, Q., Jia, C.: {PackGenome}: {Automatically} {Generating} {Robust} {YARA} {Rules} for {Accurate} {Malware} {Packer} {Detection}. In: Proceedings of the 2023 {ACM} {SIGSAC} {Conference} on {Computer} and {Communications} {Security}. pp. 3078--3092. {CCS} '23, Association for Computing Machinery (2023). \doi{10.1145/3576915.3616625}

\bibitem{li_consistently-executing_2019}
Li, X., Shan, Z., Liu, F., Chen, Y., Hou, Y.: A {Consistently}-{Executing} {Graph}-{Based} {Approach} for {Malware} {Packer} {Identification}. IEEE Access  \textbf{7},  51620--51629 (2019). \doi{10.1109/ACCESS.2019.2910268}

\bibitem{li_graph_2019}
Li, Y., Gu, C., Dullien, T., Vinyals, O., Kohli, P.: Graph {Matching} {Networks} for {Learning} the {Similarity} of {Graph} {Structured} {Objects}. In: Chaudhuri, K., Salakhutdinov, R. (eds.) Proceedings of the 36th {International} {Conference} on {Machine} {Learning}. Proceedings of {Machine} {Learning} {Research}, vol.~97, pp. 3835--3845. PMLR (Jun 2019)

\bibitem{liu_2-spiff_2021}
Liu, H., Guo, C., Cui, Y., Shen, G., Ping, Y.: 2-{SPIFF}: a 2-stage packer identification method based on function call graph and file attributes. Applied Intelligence  \textbf{51}(12),  9038--9053 (2021). \doi{10.1007/s10489-021-02347-w}

\bibitem{segdroid}
Liu, Z., Wang, R., Japkowicz, N., Gomes, H.M., Peng, B., Zhang, W.: Segdroid: An android malware detection method based on sensitive function call graph learning. Expert Syst. Appl.  \textbf{235}(C) (Jan 2024), \url{https://doi.org/10.1016/j.eswa.2023.121125}

\bibitem{intelpin}
Luk, C.K., Cohn, R., Muth, R., Patil, H., Klauser, A., Lowney, G., Wallace, S., Reddi, V.J., Hazelwood, K.: Pin: building customized program analysis tools with dynamic instrumentation. SIGPLAN Not.  \textbf{40}(6),  190–200 (jun 2005). \doi{10.1145/1064978.1065034}

\bibitem{low_entropy_analysis}
Mantovani, A., Aonzo, S., Ugarte-Pedrero, X., Merlo, A., Balzarotti, D.: Prevalence and impact of low-entropy packing schemes in the malware ecosystem. Proceedings 2020 Network and Distributed System Security Symposium  (2020)

\bibitem{muchnick_advanced_1997}
Muchnick, S.S.: Advanced {Compiler} {Design} and {Implementation}. Morgan Kaufmann, San Francisco, CA (1997)

\bibitem{file_packing_survey}
Muralidharan, T., Cohen, A., Gerson, N., Nissim, N.: File packing from the malware perspective: Techniques, analysis approaches, and directions for enhancements. ACM Comput. Surv.  \textbf{55}(5) (dec 2022). \doi{10.1145/3530810}

\bibitem{winlicense}
{Oreans Technologies}: Winlicense. \url{https://www.oreans.com/WinLicense.php}, accessed: 2024-07-08

\bibitem{sokdisassemblers}
Pang, C., Yu, R., Chen, Y., Koskinen, E., Portokalidis, G., Mao, B., Xu, J.: Sok: All you ever wanted to know about x86/x64 binary disassembly but were afraid to ask. In: 2021 IEEE Symposium on Security and Privacy (SP). pp. 833--851 (2021)

\bibitem{peid}
PEiD: Peid. \url{https://www.aldeid.com/wiki/PEiD} (2024), accessed: 2024-04-15

\bibitem{detection_packed_exes}
Perdisci, R., Lanzi, A., Lee, W.: Classification of packed executables for accurate computer virus detection. Pattern Recognition Letters  \textbf{29}(14),  1941--1946 (2008). \doi{10.1016/j.patrec.2008.06.016}

\bibitem{team_radare2_2023}
radare2: radare2: {Unix}-like reverse engineering framework and command-line tools (2024), \url{https://github.com/radareorg/radare2}, accessed: 2024-04-15

\bibitem{autoyara}
Raff, E., Zak, R., Lopez~Munoz, G., Fleming, W., Anderson, H.S., Filar, B., Nicholas, C., Holt, J.: Automatic yara rule generation using biclustering. In: Proceedings of the 13th ACM Workshop on Artificial Intelligence and Security. p. 71–82. AISec'20, Association for Computing Machinery, New York, NY, USA (2020). \doi{10.1145/3411508.3421372}

\bibitem{rahbarinia_exploring_2017}
Rahbarinia, B., Balduzzi, M., Perdisci, R.: Exploring the {Long} {Tail} of ({Malicious}) {Software} {Downloads}. In: 2017 47th {Annual} {IEEE}/{IFIP} {International} {Conference} on {Dependable} {Systems} and {Networks} ({DSN}). pp. 391--402 (2017). \doi{10.1109/DSN.2017.19}

\bibitem{virtbasedpackers}
Rolles, R.: Unpacking virtualization obfuscators. In: Proceedings of the 3rd USENIX Conference on Offensive Technologies. p.~1. WOOT'09, USENIX Association, USA (2009)

\bibitem{sun_pattern_2010}
Sun, L., Versteeg, S., Bozta{\c{s}}, S., Yann, T.: Pattern recognition techniques for the classification of malware packers. In: Steinfeld, R., Hawkes, P. (eds.) Information Security and Privacy. pp. 370--390. Springer Berlin Heidelberg, Berlin, Heidelberg (2010)

\bibitem{ugarte-pedrero_sok_2015}
Ugarte-Pedrero, X., Balzarotti, D., Santos, I., Bringas, P.G.: {SoK}: {Deep} {Packer} {Inspection}: {A} {Longitudinal} {Study} of the {Complexity} of {Run}-{Time} {Packers}. In: 2015 {IEEE} {Symposium} on {Security} and {Privacy}. pp. 659--673 (2015). \doi{10.1109/SP.2015.46}

\bibitem{dailymalware}
Ugarte-Pedrero, X., Graziano, M., Balzarotti, D.: A close look at a daily dataset of malware samples  \textbf{22}(1) (Jan 2019)

\bibitem{upx}
UPX: {UPX} -- the ultimate packer for executables. \url{https://upx.github.io/} (2024), accessed: 2024-04-15

\bibitem{symbolic_execution}
Yadegari, B., Debray, S.: Symbolic execution of obfuscated code. In: Proceedings of the 22nd ACM SIGSAC Conference on Computer and Communications Security. p. 732–744. CCS '15, Association for Computing Machinery, New York, NY, USA (2015). \doi{10.1145/2810103.2813663}

\bibitem{zaki_data_2020}
Zaki, M.J., Meira~Jr, W.: Data mining and machine learning: fundamental concepts and algorithms. Cambridge University Press, 2 edn. (2020)

\end{thebibliography}

\end{document}